\documentclass[%
 reprint,
 bibnotes,
 amsmath,amssymb,
 aps,
 pra,
]{revtex4-2}

\usepackage{graphicx}
\usepackage{dcolumn}
\usepackage{bm}
\usepackage{hyperref}
\usepackage{xcolor}

\begin{document}


\title{Coupling of magnetic and optomechanical structuring in cold atoms}

\author{T. Ackemann$^1$\email[Corresponding author: ]{ thorsten.ackemann@strath.ac.uk}, G. Labeyrie$^2$, A. Costa Boquete$^1$, G. Baio$^1$, J. G. M. Walker$^1$,  R. Kaiser$^2$, G.-L. Oppo$^1$, G. R. M. Robb$^1$ }

\affiliation{$^1$
 SUPA and Department of Physics, University of Strathclyde, Glasgow G4 0NG, Scotland, United Kingdom
}
\affiliation{$^2$Universit\'{e} C\^{o}te d'Azur, CNRS, Institut de Physique de Nice, 06560 Valbonne, France}

\date{\today}

\begin{abstract}
Self-organized phases in cold atoms as a result of light-mediated interactions can be induced by coupling to internal or external degrees of the atoms. There has been growing interest in the interaction of internal spin degrees of freedom with the optomechanical dynamics of the external centre-of-mass motion. We present a  model for the coupling between magnetic and optomechanical structuring in a $J=1/2 \to J'=3/2$ system in a single-mirror feedback scheme, being representative for a larger class of diffractively coupled systems such as longitudinally pumped cavities and counter-propagating beam schemes. For negative detunings, a linear stability analysis demonstrates that optical pumping and optomechanical driving cooperate to create magnetic ordering. However, for long-period transmission gratings the magnetic driving will strongly dominate the optomechanical driving, unless one operates very close to the existence range of the magnetic instability. At small lattice periods, in particular at wavelength-scale periods, the optomechanical driving will dominate.
\end{abstract}

\maketitle


\newcommand{\rp}{\mathbf{r}}
\newcommand{\fdip}{\mathbf{f}_{\textrm{dip}}}
\newcommand{\nabp}{\nabla_{\perp}}

\section{Introduction}
Recent years have seen interest in self-organization in cold atoms in transversely pumped cavities \cite{domokos02,black03,dimer07,baumann10,ritsch13,zhiqiang17,kollar17,landini18,guo21a,kirton19,mivehvar21a} and via diffractive coupling in longitudinally pumped cavities \cite{tesio12}, counter-propagating beam schemes \cite{muradyan05,greenberg11,schmittberger16,schmittberger16a} and single-mirror feedback schemes \cite{labeyrie14,robb15,kresic18,labeyrie18,zhang18}. In these schemes coupled light-matter structures are created via optical nonlinearities and the back action of the structured matter on light. In this manuscript, we concentrate on single-mirror feedback schemes, but we anticipate that the main conclusions are also valid for other systems. In the single-mirror feedback system, effective coupling between atoms is provided by retroreflecting back a laser beam which initially interacted with the atoms. Any spatial structure within the atomic sample will influence the refractive index of the cloud and imprint onto the phase of the transmitted light. By diffractive dephasing in the feedback loop the retroreflected light will acquire a corresponding structure which will sustain the atomic structure (see \cite{ackemann21} for a recent review).  The spontaneous formation of atomic density patterns due to optomechanically mediated interaction was demonstrated in \cite{labeyrie14}, where the light-matter interaction is provided via the dipole force. Spontaneous magnetic ordering of dipolar and quadrupolar nature was demonstrated in \cite{kresic18,labeyrie18,kresic19}, where light-matter interaction is mediated by optical pumping. The interplay between optomechanical structures and the electronic two-level nonlinearity was studied in \cite{labeyrie14,camara15}. In this contribution, we are addressing the question whether optomechanical and magnetic ordering can exist together and potentially support each other. This complements recent interest in mixed spin-density textures in cavity QED systems  \cite{mivehvar17,jaeger17,kohler18,kroeze18,landini18,masalaeva21,mivehvar21a}.

\begin{figure}[h]	
\includegraphics[width=\columnwidth]{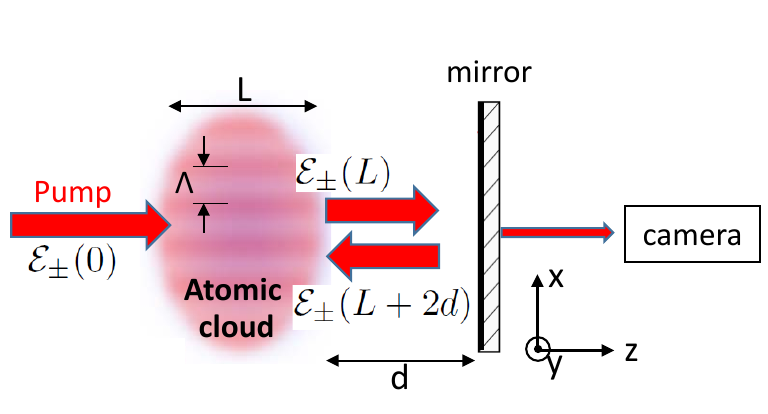}
\caption{ \label{fig:setup}
Experimental scheme for single mirror feedback. A cloud of length $L$ is driven by pump laser beam. A plane mirror at a distance $d$ retro-reflects the transmitted light back into the cloud. The theoretical treatment assumes $d>> L$. The optical axis and quantization axis is the $z$-direction. Structuring is depicted here in the $x$-direction but could by symmetry be anywhere in the transverse $x-y$-plane. See text for further explanation.
}
\end{figure}

\begin{figure}[h]	
\includegraphics[width=\columnwidth]{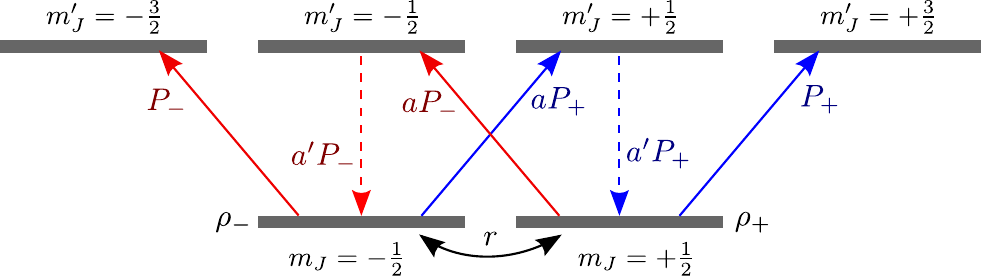}
\caption{ \label{fig:line}
$J=1/2 \to J'=3/2$ transition. $a=1/3$ denotes the relative strength of weak and strong transition. $a'=2/9<a$ allows to account for a lower optical pumping efficiency as some cycles will return to original state. The main part of the dynamics takes place in the stretched states with maximum modulus of the magnetic quantum number.
}
\end{figure}

\begin{figure}[h]	
\includegraphics[width=\columnwidth]{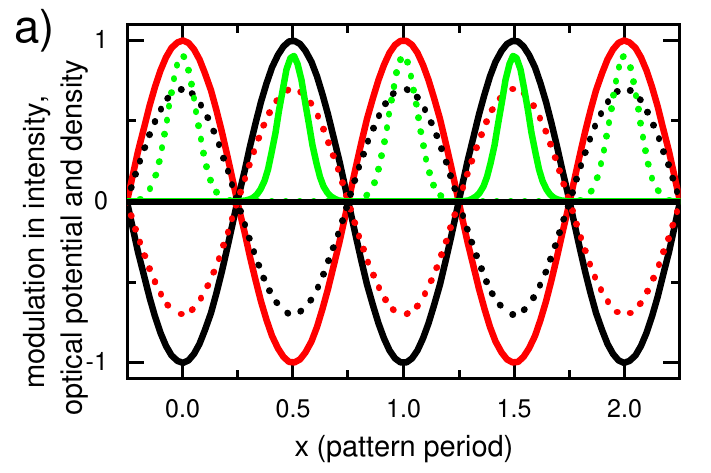}\\
\includegraphics[width=\columnwidth]{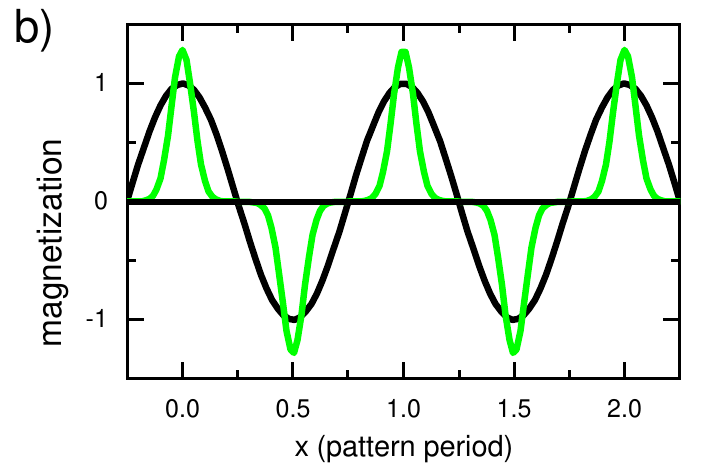}
\caption{ \label{fig:lightshift}Schematic illustration of interplay of magnetic and optomechanical degrees of freedom. The figures depict atomic states and optical fields and potentials in transverse space $x$ measured in units of the lattice period $\Lambda$. Intensity of $\sigma_+$ (solid red, dark grey in print) and $\sigma_-$ (solid black) component of optical field. Dotted red (dark grey in print) and black lines are the corresponding potentials for spin-up (dotted red, dark grey in print) and spin-down (dotted black) atoms. The green (light grey in print) lines indicate the resulting bunching in density spatially anti-phased for spin-up (dotted green/light grey) and spin-down (solid green/light grey) atoms. The atomic density is narrower than the intensity for cold (enough) atoms. b) Black line: Atomic magnetization (dominance of spin-up atoms where positive) resulting from optical pumping from the optical spin patterns (solid red and black lines in a) at low saturation. Green (light grey in print) line: more peaked structure of magnetization if atoms bunch in addition. The profiles of the curves are artist impressions and not calculated self-consistently.
}
\end{figure}

A simplified $J=1/2 \to J’=3/2$ model for the Rb transition (Fig.~\ref{fig:line}) allows us to take into account dipolar magnetic structures analogous to spin-1/2 magnets and the optomechanical effects due to light-shifts. Fig. \ref{fig:lightshift}a illustrates the magnetization of atoms and the light spin patterns on the one hand, and the dipole potentials and the resulting bunching on the other hand. It confirms that for negative detuning the two mechanisms will support each other: Where optical pumping by the optical spin structures leads to a prevalence of, e.g., population with a positive magnetic quantum number (red line and green dotted line in Fig.~\ref{fig:lightshift}a at, e.g., point 0) this state will have also the lower energy in the dipole potential (dotted red line at point 0). Hence, one expects that thresholds are lower and that the resulting magnetization peaks are narrower, more “spiky”, than without the bunching effect (Fig. \ref{fig:lightshift}b). The latter might have interesting consequences for the interaction range on the self-induced lattice as discussed in \cite{zhang18,ackemann21}. Ref.~\cite{schmittberger16a} discusses the optomechanical part in a $J=1/2 \to J’=3/2$ model following a wave-mixing approach. However, it neglects the intrinsic magnetic instability, i.e.\ would not lead to an instability for stationary atoms. We will develop a model which can describe both situations in their limiting cases, but will use the simplest possible optomechanical model for this first assessment, overdamped dynamics in an externally imposed molasses as originally suggested by \cite{muradyan05} for diffractively mediated transverse self-organization. The experiments on optomechanical self-organization reported in \cite{labeyrie14} did not use optical molasses and were interpreted in the framework of a conservative Vlasov-equation model \cite{labeyrie14,tesio14}. Optical molasses, potentially at lower amplitude and different detuning than used for the initial trapping,  are also interesting for extending the lifetime of the structured state against the heating from scattered pump photons, as earlier demonstrated for the CARL (collective atomic recoil lasing) instability \cite{voncube04}. First indications for the extension of lifetime of structured transverse states were found in counter-propagating beams \cite{schmittberger16a} and single-mirror feedback schemes \cite{gomes15u}. Most importantly for our present intentions, the simplicity of the overdamped model compared to the Vlasov model makes it suited for a first assessment of the interaction between optomechanical and magnetic ordering. As the overdamped and conservative model have the same threshold condition for the same initial temperature
\cite{tesio14,tesio14t,baio21}, the results developed here for the combined magnetic-overdamped optomechanical model will also provide a guidance for the combined magnetic - conservative optomechanical situation. Hence, the current investigations are fruitful even if scattering of molasses photons can be expected to be detrimental to the magnetic ordering as it will provide a background of photons with all polarizations.

\section{Model}

In the experimental scheme (Fig.~\ref{fig:setup}) a cold atomic cloud of $^{87}$Rb atoms is driven by a detuned lasers beam close to the $D_2$-line. The transmitted beam is phase modulated by any structure (with transverse spatial period $\Lambda$) in the gas which could be due to a magnetic ordering of Zeeman sub-states or a density modulation due to atomic bunching. The beam is retro-reflected by a plane mirror of reflectivity $R$. Diffractive dephasing leads to an amplitude modulation of the backward beam which can then sustain the structure in the cloud. Details can be found in \cite{labeyrie14,kresic18,labeyrie18,ackemann21}. Self-organization via single-mirror feedback was originally predicted in \cite{firth90a}.

For the description of the internal degrees of freedom we follow the spirit of the approach of \cite{mitschke86} for a $J=1/2\to J'=1/2$ transition, adapted to a $J=1 \to J'=2$-transition in \cite{kresic18,labeyrie18}. In these works, the dynamics for the ground state magnetization is derived in a semi-classical approximation from the Liouville equation for the density matrix. All excited state populations, excited state coherences and the optical coherence (dipole moment) are adiabatically eliminated as they evolve on the time scale of the lifetime of the excited state (i.e. $< 100$ ns) whereas the dynamics of the ground state evolves on hundreds of microseconds (see \cite{kresic18,labeyrie18} and below). Afterwards it is assumed that the total population resides in the ground state. As we are interested in a first assessment of the interplay of internal (magnetic) and external (optomechanical degrees of freedom), we consider a further simplification by neglecting excited state coherences altogether and just consider rate equations for atomic populations (as used in the discussion of Fig.~\ref{fig:line}). In principle, one could have constructed an even simpler rate equation model based on a $J=1/2 \to J’=1/2$-transition which would have avoided the complication of multiple pump paths (i.e.\ ~$a P_\pm$, $a'P_\pm$) and potential excited state coherences altogether but we decided to go for the $J=1/2 \to J’=3/2$-transition as it is keeps the experimental feature typical for the Rb D$_2$-line that the final state of optical pumping is bright and not dark. The model will be valid in the regime where the pump rate is much lower than the decay rate $\Gamma$ of the excited state populations and coherences, i.e.\ saturation values of about a few times 0.01 to about 0.1, depending on the level of qualitative or quantitative accuracy one is interested in.

The optomechnical model is based on overdamped dynamics by means of an additional external optical molasses as used in \cite{muradyan05,baio21}. Under overdamped conditions, the atom density $\rho(\rp,t)$ obeys a Smoluchowski drift-diffusion equation \cite{ohara01,ritsch13}
\begin{eqnarray} \label{eq:smolu1}
\partial_t \rho(\rp,t) 
& = & \frac{2\sigma D}{\Gamma}\, \nabp\cdot\left[\rho(\rp,t)\,\nabp P(\rp,t)\right] \nonumber \\ && +D\nabp^{2}\rho(\rp,t) 
\label{eq:smolu3}
\\ & = & \frac{2\sigma D}{\Gamma}\, \left[ \nabp \rho \nabp P + \rho \nabp^{2} P \right] +D\nabp^{2}\rho(\rp,t)
\label{eq:smolu4}
\end{eqnarray}
where $D$ is the cloud diffusivity, $P(\rp,t) = \Gamma s/2$ the pump rate proportional to the total light intensity and the saturation parameter $s(\rp,t)$. $\Delta$ corresponds to the light-atom detuning in units of the full linewidth $\Gamma$. The relative strength of optomechanical driving to thermal fluctuations is characterized by
\begin{equation}
\sigma = \frac{\hbar\Gamma\Delta}{2k_{B}T},
\end{equation}
where $k_B$ represents the Boltzmann constant and $T$ the temperature of the cloud. The formulation of Eq.~(\ref{eq:smolu4}) is beneficial as the grad-grad term can be neglected in the linear stability analysis (LSA) as it is second order in spatially inhomogeneous perturbations. It should be noted that $\sigma \sim 1/T$ and $D \sim T$ (see Eq.~(\ref{eq:D}) below), hence the driving of the optomechanical instability is actually independent of temperature but the counteracting fluctuations increase with temperature. As in a dilute thermal gas with well controlled stray magnetic fields there is no relaxation mechanism for magnetization structures other than the residual atomic motion, this implies that the temperature dependence of the optomechanical and magnetic instabilities will be similar.

The treatment of the magnetization is based on the $J=1/2 \to J'=3/2$ depicted in Fig.~\ref{fig:line}. The quantization axis is chosen to be the wavevector of the pump light. Under the assumption discussed before that we are interested in rate equations for the ground state populations only, the equations of motions are for the populations $\rho_\pm$ for the ground states with $m_j=\pm 1/2$ are
\begin{eqnarray}
  \dot{\rho_+}&=& -\frac{r}{2}(\rho_+ -\rho_- ) + a'P_+\rho_- - a'P_-\rho_+ \, + \,D\nabp^{2}\rho_+  \\ && + \frac{2\sigma D}{\Gamma} \left[ \nabp \rho_+ \nabp (P_+ +a P_-) + \rho_+ \nabp^{2} (P_+ +a P_-) \right]  \nonumber\\
  \dot{\rho_-}&=& -\frac{r}{2}(\rho_- -\rho_+ ) - a'P_+\rho_- + a'P_-\rho_+ \,  +D\nabp^{2}\rho_- \\ && + \frac{2\sigma D}{\Gamma}\, \left[ \nabp \rho_- \nabp (P_- +a P_+)+ \rho_- \nabp^{2} (P_- +a P_+)\right] \nonumber
\end{eqnarray} 
with $a=1/3$ and $a'= 2/9$ \cite{dalibard89}.

The effective decay rate $r$  for the magnetization due to the residual atomic motion introduced in \cite{kresic18} is
\begin{equation}\label{eq:r_T}
r = \frac{4}{\pi \Lambda}\,\sqrt{\frac{8k_B T}{\pi M}} .
\end{equation}
It results from an average time of atoms needing to cross ballistically a pattern period if no velocity damping by a molasses is present. It will be dropped in the combined model with diffusive damping by $D$. However, it is useful to keep for the moment to compare to the established theory for magnetic self-organization.

The populations are separated into total density $\rho$ and orientation $w$
\begin{eqnarray}
  \rho &=& \rho_+ +\rho_- \\
  w & =& \rho_+ -\rho_- \\
  \rho_+ &=& \frac{w+\rho}{2} \\
  \rho_- &=& \frac{-w+\rho}{2} . 
\end{eqnarray}

To make connection to the 2-level optomechanical model,  $\phi_S$ is the linear phase shift if all population is in one of the stretched states. Then the linear phase shift for equal population $\rho_+=\rho_-=1/2$ is
\begin{equation} \phi_{lin}=\frac{1+a}{2}\,\phi_S =  b_0\,\frac{ \Delta}{ \Gamma\,(1+(2\Delta /\Gamma )^2)}, \end{equation} where $b_0$ is the optical density in line centre measured for equal Zeeman populations. This formulation implies the normalization $\rho=1$ in the homogeneous state. The equations of motion are \begin{widetext}
\begin{eqnarray}
 \label{eq:PDE}
  \dot{w} &=& - rw  +a' (P_+ +-P_-)\rho - a'(P_++P_-)w \\
  && +D\nabp^{2} w\,+\, \frac{\sigma D}{\Gamma}\, \left[w(1+a)\nabp^{2}(P_+ + P_-)+ \rho (1-a)\nabp^{2}(P_+  -P_-)\right]\\
 && +\frac{\sigma D}{\Gamma}\, \nabp (w+\rho )\,\cdot\, \nabp (P_+ + aP_-)   +\frac{\sigma D}{\Gamma}\, \nabp (w-\rho )\,\cdot\, \nabp (P_- + aP_+) \\
  \dot{\rho} &=& D\nabp^{2} \rho +  \frac{\sigma D}{\Gamma}\,\left[\rho(1+a)\nabp^{2}(P_+ + P_-)+ w (1-a)\nabp^{2}(P_+  -P_-)\right]\\
 && +\frac{\sigma D}{\Gamma}\, \nabp (w+\rho )\,\cdot\, \nabp (P_+ + aP_-)   +\frac{\sigma D}{\Gamma}\, \nabp (w-\rho )\,\cdot\, \nabp (P_- + aP_+)  .
\end{eqnarray} \end{widetext}

The transmitted field is given by
\begin{eqnarray}
\mathcal{E}_{\pm}(L) & = &  \mathcal{E}_{\pm}(0) \exp{(i \phi_S\rho_\pm + ia \phi_S\rho_\mp)}\\
 & = &  \mathcal{E}_{\pm}(0) \exp{(i \phi_{lin}\rho \pm i \phi_{lin} \frac{1-a}{1+a} \, w)} \end{eqnarray}
where $\mathcal{E}_{\pm}$ refers to the complex field amplitude of the $\sigma_\pm$-polarization component which are scaled such that their squares are the pump rates $P_\pm$. The argument in brackets refers to the position on the $z$-axis. $\mathcal{E}_{\pm}(0)$ is the input field at the entrance of the cloud and $\mathcal{E}_{\pm}(L)$ the one at the exit. It is assumed that the atomic variables do not change over $z$ (quasi-2D situation). It is also assumed that the cloud is diffractively thin, i.e.\ we can neglect diffraction within the medium.  The backward field is then obtained in Fourier space. The diffractive dephasing between the on-axis pump and the off-axis spontaneous sidebands of the field is described by the phasor $\Theta = q^2\times z/(2k)= q^2\times d/k$ after propagation of a distance $z=2d$ to the mirror and back \cite{firth90a,ackemann21}. The total pump rate is taken as the sum of the intensities of the forward and backward field neglecting interference of the counterpropagating beams and the approximately wavelength scale gratings created by it. The restriction to a diffractively thin medium and the neglect of wavelength-scaled gratings is useful for a first discussion of the interaction between optomechanical and optical pumping nonlinearities and is sufficient for a qualitative description of our experimental situation in most circumstances \cite{ackemann21}. Extensions are quite complex and discussed for the simple case of a two-level nonlinearity in \cite{firth17}. We will comment on limitations of this approach where appropriate below.

The input field is linearly polarized, $ |\mathcal{E}_{+}(0)|^2= |\mathcal{E}_{-}(0)|^2 =P_0$. For a $J=1/2$ ground state allowing only for dipolar ordering the phase between the $\sigma_\pm$-components is not important for the optical pumping in ground state. Hence any transverse input polarization is possible. The spontaneous emergence of $\pi$-light (polarized parallel to the quantization axis) is not expected and was never reported in the literature for longitudinal pumping as it would demand a coupling via a checkerboard pattern in the longitudinal and at least one transverse direction on wavelength scales. This is the natural situation in transversely pumped cavities but not in systems with a single distinguished axis as the single-mirror feedback system, counter-propagating beams and longitudinally pumped cavities.

The homogenous solution for the system is a homogenous density $\rho_h=1$ and zero orientation, $w_h=0$. The ansatz for the linear stability analysis is then
\begin{eqnarray}
  \rho &=& 1+ \delta \rho \\
  w &=& \delta w ,
\end{eqnarray}
where $\delta \rho$ and $\delta w$ are small spatially periodic functions, $\sim cos{(\vec{q}\cdot\vec{r})}$,  with wavevector $\vec{q}$ and position vector $\vec{r}$ in 2D transverse space. As the system is rotational symmetric, the threshold condition is not depending on the direction of $\vec{q}$ but only on the wavenumber $q=|\vec{q}|$ (this was already used in the calculation of the diffractive phasor $\Theta$) and can be obtained by considering a single spatial harmonic. The linear expansion for the transmitted field yields
\begin{eqnarray}
  \mathcal{E}_{\pm}(L) &\approx & \mathcal{E}_{\pm}(0) \exp{(i \phi_{lin})} \, \times \nonumber \\
  && (1+ i \phi_{lin}\delta \rho)(1\pm i \phi_{lin}\delta w \frac{1-a}{1+a} )\\
  &\approx & \mathcal{E}_{\pm}(0) \exp{(i \phi_{lin})} \, \times  \nonumber \\
  &&(1+ i \phi_{lin}\delta \rho \pm i \phi_{lin}\delta w \frac{1-a}{1+a} ).
\end{eqnarray}
This yields 
\begin{eqnarray}
  \mathcal{E}_{\pm}(L) & = & \mathcal{E}_{\pm}(0) \exp{(i \phi_{lin})} \,(1+i e^{i\Theta} \phi_{lin}\delta \rho)\, \times \nonumber \\&&
   (1\pm i e^{i\Theta}  \phi_{lin}\delta w \frac{1-a}{1+a} )\\
   |\mathcal{E}_{\pm}(L+2d)|^2 &\approx & |\mathcal{E}_{\pm}(0)|^2
    (1 - 2 \phi_{lin} \sin{\Theta} \delta \rho \, \nonumber \\
    &&\mp 2\phi_{lin}\sin{\Theta}\delta w \frac{1-a}{1+a} )\\
   P_{\pm}(L+2d) & = & P_0  \,
   (1 - 2 \phi_{lin}\sin{\Theta}\delta \rho \, \nonumber \\
   &&\mp 2\phi_{lin}\sin{\Theta}\delta w \frac{1-a}{1+a} ),
\end{eqnarray}
where it is used that the input light is  linearly polarized and hence $|\mathcal{E}_+(0)|^2 =|\mathcal{E}_-(0)|^2 =P_0$. Note that the phase of the reentrant $\sigma_\pm$-fields is not affected for a $J=1/2$ ground state, i.e.\ the polarization direction is not modulated, just the helicity.

Linear stability analysis for the density gives
\begin{equation}
\delta\dot{\rho} = -Dq^2 \delta \rho
 + \frac{4\sigma D \phi_{lin} \sin{\Theta} P_0 R q^2}{\Gamma} \, (1+a) \, \delta \rho
 \end{equation}
 with threshold
\begin{equation}\label{eq:optothres}
 P_{0,th} = \frac{\Gamma}{4\phi_{lin} \sin{\Theta}\sigma R (1+a)} .
\end{equation}
Note that for linearly polarized light $P_{0,th}$ represents only half the input intensity. This results agrees with the expression for the scalar case in \cite{baio21} for $a=0$. Minimal threshold is achieved for $\sin{\Theta}=1$ which corresponds to a phase shift of $\pi/2$ after a quarter of the Talbot distance. This provides positive feedback for a self-focusing situation \cite{firth90a,labeyrie14,ackemann21}. Note that this is independent of the sign of detuning as both $\sigma$ and $\phi_{lin}$ change sign with detuning. The threshold condition for the overdamped case is the same as for the conservative case at the same initial temperature. Eq.~(6) of \cite{tesio14} reduces to Eq.~(\ref{eq:optothres}) for high enough optical densities. At these optical densities (and the one considered below) the optical dipole potential can be taken as being proportional to $\log{(1+P)}\approx P$. Using  $\log{(1+P)}$ as the expression for the dipole potential in the derivation here, would lead to Eq.~(6) of \cite{tesio14}.

The LSA for the orientation gives
\begin{eqnarray}
\label{eq:lsa_w}
  \delta\dot{w} & = & -r \delta w - Dq^2 \delta w - 2a'P_0(1+R) \delta w \nonumber\\
  && -4RP_0\phi_{lin} \sin{\Theta} a' \frac{1-a}{1+a}\, \delta w\\
 &&+ \frac{4\sigma D \phi_{lin} \sin{\Theta} P_0 R q^2}{\Gamma} \, \frac{(1-a)^2}{1+a} \, \delta w \nonumber,
\end{eqnarray}
where the first line describes decay and saturation by total pump rate, the second line driving by optical pumping and the last line driving by optomechanics. The magnetic decay and driving terms have the same structure as derived in \cite{kresic18} for a $J=1\to J'=2$-transition but with different numerical values for the pre-factors. The term describing optomechanical driving of a magnetic state in the last line is the new one derived in this treatment. For negative detuning ($\Delta <0$, $\sigma <0$, $\phi_{lin}<0$), minimal threshold is reached for $\sin{\Theta}=1$ and the optomechanical effect enhances the magnetic instability as expected from the considerations in Fig.~\ref{fig:lightshift}. For positive detuning, optimal feedback for the magnetic instability is obtained for $\sin{\Theta}=-1$ and for the optomechanical instability for $\sin{\Theta}=1$, i.e.\ the two instabilities are opposing each other. This is because the magnetic one is self-defocusing (as the trapping state is bright on the D$_2$-line the optical pumping nonlinearity is defocusing for blue detuning) and the optomechanical one is focusing. In the potential picture of Fig.~\ref{fig:lightshift}, for blue detuning the optically pumped atoms would be expelled from the intensity maxima optimally sustaining the magnetic ordering. The threshold is given by
\begin{widetext}\begin{eqnarray}
\label{eq:Pth_w}
  P_{0,th} & = \frac{Dq^2} { - 2a'(1+R) -4R\phi_{lin} \sin{\Theta} a' \frac{1-a}{1+a}\, + 4R \phi_{lin} \sin{\Theta} q^2 \, \frac{\sigma D}{\Gamma} \, \frac{(1-a)^2}{1+a}} .
\end{eqnarray}\end{widetext}


An interesting aspect is that the two instabilities are not directly coupled in linear order as the equations for $\delta \rho$ and $\delta w$ are not coupled to each other. The reason seems to be that a magnetic instability at $q$ would drive a density modulation at $2q$. A density modulation at $q$ will drive a magnetic modulation at $q/2$. These are second order processes.

\section{Analysis}

\begin{figure}[h]	
\includegraphics[width=\columnwidth]{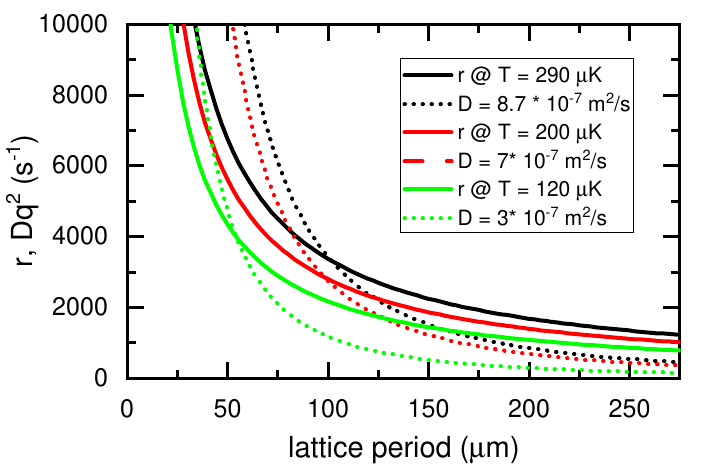}
\caption{ \label{fig:relaxation}
Effective relaxation rate (solid lines) vs.\ lattice period from Eq.~(\ref{eq:r_T}) for the temperatures relevant for \cite{labeyrie14,labeyrie18,kresic18} compared to diffusive modelling (dotted lines) for different diffusion constants. Black (corresponding to situation in \cite{labeyrie14}): solid line $T=290$~$\mu$K, dotted line $8.7\times 10^{-7}$m$^2$/s; red (dark grey in print, corresponding to situation in \cite{labeyrie18}): solid line $T=200$~$\mu$K, dotted line $7\times 10^{-7}$m$^2$/s; green (light grey in print, corresponding to situation in \cite{kresic18}): solid line $T=120$~$\mu$K, dotted line $3\times 10^{-7}$m$^2$/s.
}
\end{figure}

In a first step, Fig.~\ref{fig:relaxation} compares the relaxation of the structures assuming ballistic motion using an effective decay rate $r$,  Eq.~(\ref{eq:r_T}), with the diffusive ansatz with a molasses (giving a decay rate of $Dq^2$) for the magnetic self-ordering. Obviously, in both cases relaxation decreases with increasing lattice period $\Lambda$, but the functional relationship is proportional to $q^{-1}$ for the ballistic case and  $q^{-2}$ for the diffusive one. Matching of rates for the experiment performed at INPHYNI using a large atomic cloud \cite{labeyrie14,labeyrie18} at the relevant scale of about $\Lambda \approx 100\,\mu$m gives a diffusion coefficient of about $(7 - 9)\times 10^{-7}$m$^2$/s. For the experiment performed at Strathclyde with a smaller cloud \cite{kresic18}, it is $D\approx 3\times 10^{-7}$m$^2$/s to match at $\Lambda \approx 50$~$\mu$m. These values are  compatible  with diffusion coefficients obtained in optical molasses. Measurements in a lin-perp-lin configuration in \cite{hodapp95} indicate values $D\approx (1- 2)\times 10^{-7}$m$^2$/s for the diffusion coefficient above saturation intensity, and a few times  $10^{-7}$m$^2$/s at lower intensities. This order of magnitude is fine for the current preliminary estimations. Properties of the molasses would need to be optimized anyway.

Temperature and diffusion coefficient are not independent in an optical molasses. Temperature is related to the momentum diffusion coefficient $D_p$ and the velocity damping coefficient $\alpha$ \cite{lett89,hodapp95} by
\begin{equation}\label{eq:Dp}
 T = \frac{D_p}{\alpha k_b} ,
\end{equation}
and the space diffusion coefficient is given by
\begin{equation}\label{eq:D_Dp}
 D = \frac{D_p}{\alpha^2} .
\end{equation}
Hence
\begin{equation}\label{eq:D}
 D = \frac{k_BT}{\alpha} .
\end{equation}
For the velocity damping coefficient we use the estimation of \cite{hodapp95} for a 3D lin-perp-lin molasses,
\begin{equation}\label{eq:Dp}
 \alpha \approx  \frac{3}{7}\hbar k^2\, |\Delta_M |,
\end{equation}
where $\Delta_M$ denoted the detuning of the molasses normalized to $\Gamma$. Detunings in the range of 1.35-1.9 will produce then the combination of temperatures and diffusion coefficients identified in Fig.~\ref{fig:relaxation} and taking $|\Delta_M |=1.8$ will sweep the diffusion coefficient from $3.1\times 10^{-7}$m$^2$/s at 120~$\mu$K to $7.6\times 10^{-7}$m$^2$/s at 290~$\mu$K, i.e.\ allows to cover the range for the first estimations intended here.

\begin{figure}[tbh]	
\includegraphics[width=\columnwidth]{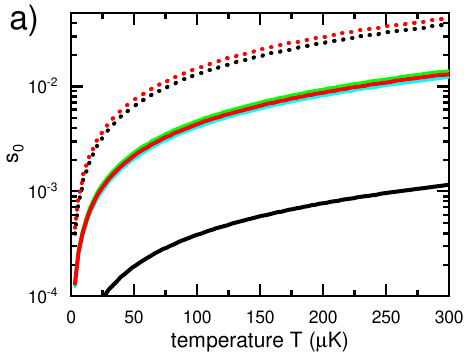}\\
\includegraphics[width=\columnwidth]{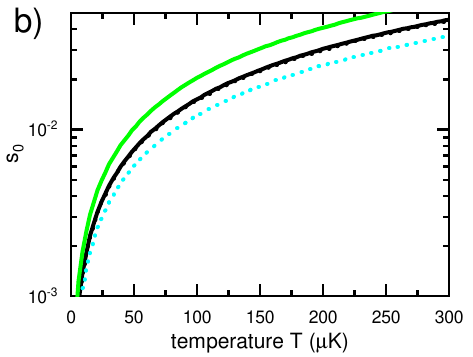}
\caption{ \label{fig:threshold_mag_opto}
Threshold saturation parameter for one circular component vs.\ temperature for $\Delta=-8.6$, $\Lambda =100$~$\mu$m. a) Black lines: $b_0=80$, solid line threshold for magnetic ordering, dotted line for optomechanical bunching, the thresholds for combined magnetic and optomechanical driving of magnetic ordering  are essentially indistinguishable from the black solid line for magnetic ordering alone at this scale. Red lines (dark grey in print): $b_0=70$, solid line threshold for magnetic ordering, dotted line for optomechanical bunching (both are above their black counterparts). The cyan solid line (light grey in print, just below the red (dark grey) solid line) represents the combined threshold for magnetic and optomechanical driving for negative detuning and $b_0=70$, the green solid line (light grey in print, just above the red (dark grey) solid line) positive detuning. b) Same for $b_0=69.31$. The cyan line is now dotted to indicate the difference to the green line in print. The parameter $\sigma$ changes from infinity at $T=0$ to 4.3 at $T=300$~$\mu$K via 12.5 at $T=100$~$\mu$K and 8.4 $T=150$~$\mu$K.}
\end{figure}

Fig.~\ref{fig:threshold_mag_opto}a shows the threshold saturation parameter, $s_0=P_0/(\Gamma/2)$, for one circular component of the linearly polarized input vs.\ temperature for the parameters analyzed in \cite{labeyrie18,kresic18}, $\Delta=-8.6$, $\Lambda =100\, \mu$m, $b_0=80$. The solid black line indicates the threshold for magnetic ordering, Eq.~(\ref{eq:Pth_w}), the dotted black line the one for optomechanical bunching, Eq.~(\ref{eq:optothres}). They are apart by almost two orders of magnitude. Indeed, the structures observed in \cite{labeyrie18,kresic18} were identified as arising from magnetic ordering. The predicted intensity thresholds rise from about 0.2~mW/cm$^2$ at 100 $\mu$K via 0.4~mW/cm$^2$ at 200 $\mu$K to 0.6~mW/cm$^2$ at 300 $\mu$K. The experimentally observed threshold for 200 $\mu$K is about 2~mW/cm$^2$. The difference is attributed to stray magnetic fields,  the finite reflectivity of the mirror and imperfect anti-reflection coatings of cell windows. The predicted intensity thresholds for the optomechanical bunching rise from about 6.4~mW/cm$^2$ at 100 $\mu$K via 13~mW/cm$^2$ at 200 $\mu$K to 19~mW/cm$^2$ at 300 $\mu$K. No experiments on optomechanics have been performed in this parameter range but the experiments in \cite{labeyrie14} can serve as a guideline as the threshold should be independent of detuning and only dependent on initial temperature at the optical densities used, independent of whether the molasses is present or not \cite{tesio14,baio21,tesio14t}. The minimum thresholds found there are about 50~mW/cm$^2$ at 290 $\mu$K, i.e.\ in reasonable agreement with the estimation allowing for some heating effects by the pump beams.


As the magnetic and optomechanical thresholds are quite different, the thresholds for combined optical pumping-optomechanical driving of the magnetic ordering are essentially indistinguishable from the optical pumping one for both detunings and not included in Fig.~\ref{fig:threshold_mag_opto}a for $b_0=80$. The reason that temperature is not a strongly discerning parameter between instabilities is that, as indicated before, the driving is independent of temperature for both instabilities ($\sigma D$ is independent of temperature) but both are counteracted  by the residual atomic motion. By introducing a transverse magnetic one can open a relaxation channel for the orientation and destroy it (e.g. \cite{kresic18}) leaving the possibility of optomechanical and/or electronic structuring. However, no detailed experimental investigations of the transition has been done.

This changes somewhat at lower optical densities as the presence of the saturation term in the expression for the magnetic ordering implies a minimum value of the linear phase shift, $\phi_{lin} > (1+R)/R$. For $\Delta=-8.6$ this implies $b_0>69.3$. The solid and dotted red (dark grey in print) lines in Fig.~\ref{fig:threshold_mag_opto}a denote the situation for $b_0=70$. Both the optomechanical and the magnetic threshold increased but the gap between them narrowed as the magnetic threshold is much more susceptible to a density change due to the saturation term. In this situation the thresholds from combined driving are discernable from the optical pumping one, leading to a reduction of threshold for negative detuning (cyan line, light grey in print) and an increase of threshold for positive detuning (green line, light grey in print). Going closer to the magnetic threshold, $b_0=69.31$, the optomechanical and magnetic thresholds nearly coincide (Fig.~\ref{fig:threshold_mag_opto}b). The cooperation between the optomechanical and magnetic driving leads now to a substantial reduction of threshold (dotted cyan line, light grey in print) for negative detuning, whereas the combined threshold is enhanced for positive detuning (green line, light grey in print) demonstrating the interaction between the driving terms. Experiments in the apparatus at Strathclyde are performed at lower optical density measured to be $b_0\approx 27$, which is very close to the minimum density required \cite{kresic18,ackemann21}. More robust agreement between experiment and theory has been obtained assuming $b_0\approx 30$ \cite{kresic19}. The current considerations indicate that the self-organization might have been helped by the reenforcement between optomechanical and magnetic degrees of freedom already under this situation. (There is no quantitative correspondence between thresholds quoted here and in \cite{kresic18,ackemann21} because the latter papers assume a more realistic $J=1\to J'=2$-transition which changes the pre-factors of order one in the susceptibilities and pump rates.) One can also note that there is a strong preference for magnetic patterns to appear for negative detuning \cite{kresic18,labeyrie18,ackemann21} and only some indications are found for positive detuning \cite{labeyrie20u}. Although this observation agrees in tendency with the prediction above, the main reason for this asymmetry is probably the fact that under self-defocusing conditions (as indicated, the optical pumping nonlinearity behaves self-defocusing for positive detuning on the D$_2$-line) the assumption of a homogenous distribution of orientation along the beam axis is not well justified, as earlier identified for defocusing situations \cite{labeyrie14,ackemann21,aumann02}.

\begin{figure}[tbh]	
\includegraphics[width=\columnwidth]{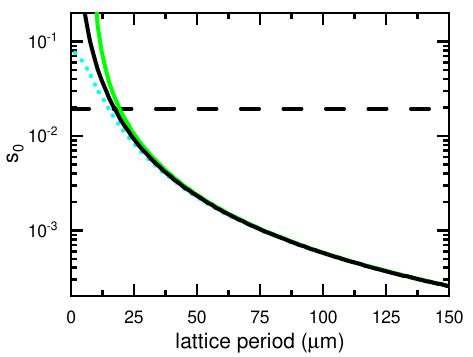}
\caption{ \label{fig:threshold_mag_period}
Threshold saturation parameter for one circular component vs.\ lattice period for $\Delta=-8.6$, $b_0=80$, $T =150\, \mu$K. Solid black line threshold for magnetic ordering, dashed black line for optomechanical bunching,  cyan (light grey in print) dotted line combined threshold for magnetic and optomechanical driving of magnetic ordering for negative detuning and green (light grey in print) solid line  combined threshold for positive detuning $\Delta=+8.6$.}
\end{figure}

However, one difference between the magnetic and optomechanical self-organization is that in the latter not only relaxation but also driving increases with increasing transverse wavenumber, i.e.\ decreasing lattice period, as the dipole force depends on the gradient of the optical potential. Hence these effects cancel and threshold is independent of wavenumber, as apparent from Eq.~(\ref{eq:optothres}). The same holds for the conservative model discussed in \cite{robb15}. In reality we expect stochastic effects due to the scattering of pump photons to favour longer period grating over shorter period ones, in particular for the conservative case, as argued in \cite{labeyrie14}. We will discuss some aspects of this further below. Fig.~\ref{fig:threshold_mag_period} shows an analysis for a representative temperature around the Doppler temperature, $T =150\, \mu$K. The horizontal dashed line indicates the optomechanical threshold for density bunching being independent of lattice period in the framework of this treatment. The magnetic threshold increases with decreasing lattice period (solid black line) as the structure is washed out by the residual transverse atomic motion. The intersection is at  $\Lambda \approx 17\, \mu$m. For smaller lattice periods the density modulation is the primary driver. In the vicinity of this point, there is again the splitting of threshold condition for the magnetic case apparent as the optomechanical driving is either supporting (dotted cyan line, $\Delta < 0$) or inhibiting the magnetic ordering (solid green line, $\Delta > 0$). It should be noted that this point is not accessible in an experiment with thermal atoms as the small transverse period violates the assumption of a diffractively thin medium. The minimum length scale which can be obtained is of the order of $\Lambda_{min} \approx \sqrt{\lambda L}$ where $L$ is the longitudinal extension of the cloud \cite{firth17,ackemann21}. This scale agrees roughly with the scale expected for an instability in two independent counter-propagating pump beams \cite{firth90,firth90c}. This implies $L_{max}\approx 0.37$~mm which is much smaller than thermal clouds with sufficient optical density to reach threshold ($L\approx 2-10$~mm in our experiments). The region in question gives also a relatively high threshold of $s\approx 0.1$ where the limitation to ground state dynamics becomes questionable. Hence, to maintain within the assumptions of the ground-state model,  the temperature of the cloud needs to be lower and/or the optical density needs to be higher demanding further optimization of the trapping and cooling process. However, these considerations can be important for extending the treatment in \cite{robb15,zhang18} on optomechanical self-organization in quantum degenerate gases with optical feedback to spinor-BECs as these have sufficient optical density in small clouds.

This limitation on period holds for gratings stemming from the interference between a pump beams and (nearly) co-propagating sidebands (so-called `transmission gratings'). Interference between a pump and (nearly) counter-propagating sidebands results in gratings with a wavenumber slightly smaller than $2k$ (so-called `reflection gratings'). Although not directly included in our analysis, the argument can be made that these gratings are mainly driven by optomechanical effects. This was indeed argued in \cite{greenberg11,schmittberger16,schmittberger16a} which treat optomechanical bunching concentrating on reflection gratings in a $J=1/2 \to J'=3/2$ transition without considering the option of a purely magnetic ordering.

\begin{figure}[tbh]	
\includegraphics[width=\columnwidth]{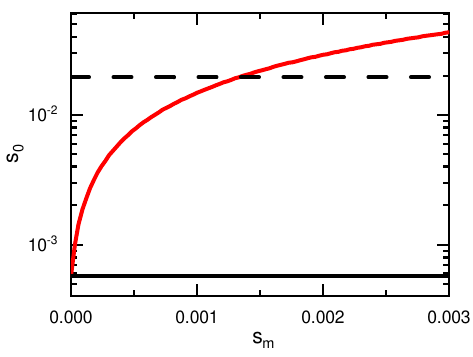}
\caption{ \label{fig:threshold_mag_molases}
Threshold saturation parameter for one circular component vs.\ saturation parameter for one molasses beams for $\Delta=-8.6$, $b_0=80$, $T =150\, \mu$K, $\Lambda =100\,\mu$m. Horizontal solid black line threshold for magnetic ordering without molasses, horizontal dashed black line for optomechanical bunching, and red line (dark grey in print) the threshold for magnetic ordering including the repumping of the molasses.}
\end{figure}

A final consideration to be done is that the presence of the 3D molasses implies a background field with many polarization components which can be expected to scramble the magnetic substates and to counteract the optical pumping by the pump beams. Formally, it can be seen also from the expression in Eq.~(\ref{eq:lsa_w}) where the intensity dependent damping term in the first row represents the action of the sum pump rate. We model the action of the molasses by adding a corresponding damping $-6P_m$ where $P_m$ is the pump rate of a single molasses beam and no allowance is made for the difference in Clebsch-Gordon coefficients between substates. We also take temperature and diffusion constant as constant for simplicity. Fig.~\ref{fig:threshold_mag_molases} illustrates that already a saturation parameter of slightly larger than $10^{-3}$ is sufficient to drive the magnetic threshold above the optomechanical one. At this point, the total saturation by pump and molasses which needs to be overcome by the driving is of the order of the optomechanical threshold. Hence, experiments on magnetic ordering will need to be performed without molasses, as in \cite{kresic18,labeyrie18,kresic19}.

 \section{Conclusions}
 We demonstrated by linear stability analysis that optical pumping and optomechanical driving will cooperate in magnetic ordering for negative detuning, in principle. However, for long period transmission gratings the driving by optical pumping will strongly dominate. Close to the existence limit of magnetically ordered states (i.e.\ at low optical density), the additional driving by optomechanics is more important and will help to sustain these states over a larger parameter regime. The background photons from the molasses will scramble the magnetic ordering at quite low saturation values. Hence, an overdamped system with molasses is not attractive for these investigations but magnetic-optomechanical coupling should be realized without molasses in the conservative limit, with a view of including the potential cooling and heating effects by the pump and self-organized gratings in the long term. Ref.~\cite{schmittberger16} found indications for 3D Sisyphus also in the transverse direction after self-organization occurred in a counter-propagating beam scheme. However, as the optomechanical threshold conditions are identical for the overdamped and conservative case for the same initial temperature
 we expect that the results obtained here with molasses are indicative what to expect in the conservative case. Current procedure to establish that the structures are of magnetic origin is their sensitivity to an external transverse magnetic field \cite{kresic18,labeyrie18}. As for well controlled magnetic fields, the residual atomic motion represents the fluctuations counteracting both instabilities, temperature cannot be used to favor optomechanical ordering vs.\ magnetic ordering. The prevalence of the magnetic structures to enable optomechanical ordering has been broken by introducing a polarizer in the feedback loop to enforce the incident linear polarization state \cite{labeyrie14}. Switching between circular and linear input polarization is an alternative way to distinguish between magnetic (or mixed) self-organization and optomechanical self-organization without putting in a polarizer in the feedback loop and as such keeping all residual losses and alignments the same, if analyzing the predicted difference between magnetic and optomechanical thresholds. In the analysis, $\phi_{lin}$ needs to be replaced by $\phi_S$ and the saturation parameter needs to be low enough to avoid excitation of the excited state. This has a parallel in the experiment of Ref.~\cite{landini18} where also a change in polarization state (there polarization direction) of the pump leads to a preference of the density modulated state over the spin-modulated state. In systems, in which the spinor interaction is not implemented via the tensor susceptibility of a transition but by engineered Raman transitions between pseudo-spins  (e.g.\ \cite{dimer07,mivehvar17,zhiqiang17,kroeze18}), there is still an interesting parallel as the threshold of the superradiance combining spin and recoil effects will depend on the difference between recoil frequency and Raman detuning, i.e.\ change with sign of detuning \cite{kroeze18}.

 Reducing the structure period increases the relative importance of the optomechanical driving and at periods below about 20 $\mu$m the optomechanical density structuring will prevail as not only relaxation but also the driving by the dipole force increases. This transition might be observable using quantum degenerate gases which have the necessary optical density in small size clouds. The results also indicates that optical pumping induced wavelength scale reflection grating are not significant compared to the long-period transmission gratings. This supports the assumption in \cite{greenberg11,schmittberger16,schmittberger16a} to treat only optomechanical effects in a counter-propagating beam scheme. In conclusion, the best prospects for studying the interaction between optomechanical and magnetic driving for thermal atoms is to use cold and small clouds with structure periods on the order of some tens of micrometers.

Having established the effects of magneto-optomechanical coupling on threshold conditions, it is now interesting to look at the nonlinear stage of evolution by numerical simulations. Even if the instabilities for density and orientation decouple in linear approximation during the initial stage of the instability, this won't be the case in the saturation regime where the structures achieved significant amplitudes. Modulations of total density will appear even if the instability is driven by the orientation and vice versa. One can expect a narrowing of structures for negative detuning (Fig.~\ref{fig:lightshift}b). Via the higher harmonics involved, this will have consequences on the coupling behaviour across the self-induced lattice \cite{zhang18,ackemann21}.


\end{document}